\documentclass[letterpaper, 10pt, conference]{ieeeconf}
\IEEEoverridecommandlockouts            
\usepackage{xspace,amssymb,epsfig,syntonly}
\usepackage{epsfig,amsmath}
\usepackage{xspace,syntonly}
\usepackage{url}
\usepackage{algorithm,algorithmic}
\usepackage{mathtools}
\usepackage{cite,bbm}

\newcommand{\E}{{\mathbb{E}}}

\newcommand{\cD}{{\cal D}}

\newcommand{\cG}{{\cal G}}

\def \N {\mathcal N}
\def \E {\mathcal E}
\def \G {\mathcal G}

\def \D {\mathcal{D}}

\newcommand{\gen}{{\mathrm{\emph g}}}

\usepackage{epstopdf}

\begin{document}

\newtheorem{definition}{Definition}
\newtheorem{remark}{Remark}
\newtheorem{proposition}{Proposition}
\newtheorem{lemma}{Lemma}
\newtheorem{corollary}{Corollary}
\newtheorem{assumption}{Assumption}
\newtheorem{theorem}{Theorem}
\newcommand{\qedblack}{\hfill \ensuremath{\blacksquare}}

\title{ \LARGE \bf Engineering Inertial and Primary-frequency Response for \\ Distributed Energy Resources}
\author{Swaroop S. Guggilam, Changhong Zhao, Emiliano Dall'Anese, Yu Christine Chen, and Sairaj V. Dhople 
\thanks{S. S. Guggilam and S. V. Dhople are with the Department of Electrical and Computer Engineering, University of Minnesota, Minneapolis, MN, USA. E-mail: {\{guggi022, sdhople\}@umn.edu}. C. Zhao and E. Dall'Anese are with the National Renewable Energy Laboratory, Golden, CO, USA. E-mail: {\{changhong.zhao, emiliano.dallanese\}@nrel.gov}. Y.~C.~Chen is with the Department of Electrical and Computer Engineering, The University of British Columbia, Vancouver, British Columbia, Canada. E-mail: {chen@ece.ubc.ca}.}}
\maketitle
\begin{abstract} We propose a framework to engineer synthetic-inertia and droop-control parameters for distributed energy resources (DERs) so that the system frequency in a network composed of DERs and synchronous generators conforms to prescribed transient and steady-state performance specifications. Our approach is grounded in a second-order lumped-parameter model that captures the dynamics of synchronous generators and frequency-responsive DERs endowed with inertial and droop control. A key feature of this reduced-order model is that its parameters can be related to those of the originating higher-order dynamical model. This allows one to systematically design the DER inertial and droop-control coefficients leveraging classical frequency-domain response characteristics of second-order systems. Time-domain simulations validate the accuracy of the model-reduction method and demonstrate how DER controllers can be designed to meet steady-state-regulation and transient-performance specifications.     
\end{abstract}

\section{Introduction} \label{sec:Introduction}
Power-system operational practices across a broad temporal spectrum will need to be refashioned to acknowledge and accommodate the increased integration of distributed energy resources (DERs) and gradual displacement of fossil-fuel driven generation~\cite{Taylor-2016}. In this paper, we focus on time scales corresponding to inertial and primary-frequency response. Frequency swings immediately following large-signal generation- or load-side disturbances are conventionally addressed exclusively with synchronous-generators' mechanical inertia and their turbine governors. The increased integration of DERs brings along the challenge of maintaining system frequency with less rotational mechanical inertia, but also offers the opportunity to exercise synthetic inertial and droop control at time scales faster than that possible with synchronous generators. While it is widely recognized that DERs ought to provide frequency regulation as part of ancillary services~\cite{Sullivan-2014,Ela-2014,Gevorgian-2015}, there are---as of yet---a limited number of system-theoretic methods to engineer the frequency response in mixed DER-generator systems.

To address the problem outlined above, our approach proceeds as follows. Beginning with a detailed third-order model adopted for individual generators (capturing rotor-angle, frequency, and mechanical-power dynamics), and a second-order model adopted for aggregations of DERs (capturing DER bus-voltage angle and electrical-frequency dynamics), we develop a reduced second-order lumped-parameter model that acknowledges all frequency-responsive devices and includes system frequency and the aggregated mechanical-power-output of generators as states.  This reduced-order model retains DER synthetic-inertia and droop-control coefficients from the original model in closed form, so that these parameters can  be designed to meet transient and steady-state regulation specifications for system frequency by leveraging well-established notions pertaining to second-order systems.  While some facets of the resulting lumped-parameter model are obvious in hindsight (e.g., net damping is the sum of generator-droop, load-damping and DER-droop constants, and net inertia is the sum of generator mechanical inertia and DER synthetic inertia), the choice of time constant for the state that captures the aggregated turbine-governor dynamics is far from apparent. Leveraging insights on the spectral properties of pertinent system matrices, we outline an optimization problem to determine this time constant. Serendipitously, we find that this is only a function of individual synchronous-generator droop and turbine-governor time constants and independent of the DER control parameters. This allows us to decouple the model-reduction method from the parameter-tuning process. 

Related prior art can be broadly grouped  into power-system model reduction methods and system-theoretic efforts to design DER synthetic-inertia and droop-control coefficients. Model-reduction methods for power systems is a widely researched topic~\cite{chow2013power}. However, the dominant theme here is the application of numerical techniques such as selective modal analysis, balanced truncation, and Krylov-subspace methods to detailed power-system dynamical models, which are perfectly known \emph{a priori}. Typically, such methods rely on myriad  matrix manipulations and factorizations that challenge the development of analytical methods to relate the parameters of the reduced-order model to those of the original one. In the recent work in~\cite{Apostolopoulou-2016}, a reduced-order dynamical model for a balancing-authority area while retaining original-system parameters was developed, but the  accuracy of the reduced-order model is not  addressed. Unlike these previous works, our proposed method: i) relates the parameters of the reduced-order system to those of the original one, ii)  justifies the validity of the reduced-order model based on spectral properties, and iii) rigorously bounds the error between the reduced-order system and the original one.

Shifting focus to literature on frequency-responsive DERs, there is a wide body of work that focuses on \emph{DER-level} controller design for inertial and primary-frequency control~\cite{Ghosh-2016,Zhao-2016,Vyver-2016}. However, limited attention has been devoted to relating synthetic-inertia and droop-control coefficients to the post-disturbance \emph{system-wide} dynamic frequency response or steady-state frequency regulation. A few notable exceptions to this general observation are the efforts in~\cite{Baldick-2014,Fei-2015,Teng-2016}; however to simplify the analysis, the methods therein adopt a constant-ramp-rate model for governors. A brute-force optimization-based approach that leverages sensitivity of frequency overshoot and damping ratio to engineer inertia and damping constants is provided in~\cite{Borsche-2015}. This method is based on explicitly computing the system eigenvalues, and is not accompanied with a proof of convergence or guarantee of scalability to larger networks. Finally, we bring to attention the effort in~\cite{Poolla-2016} that addresses the tangentially related problem of determining optimal locations in the network to locate synthetic inertia.  While we assume the DER locations are fixed, once the aggregate inertial and droop-control parameters are determined, we outline how these could be optimally allocated between DERs to ensure power sharing. This builds on our previous work which focused on developing notions of participation factors for allocating DER primary-frequency response~\cite{Swaroop-2016}.  

The remainder of this manuscript is organized as follows. In Section~\ref{section:dynamics} we outline the dynamical models adopted for the generators and DERs, and in Section~\ref{sec:modelreduction}, we obtain the reduced second-order model. The approach to design the DER control coefficients and numerical simulations to validate the model reduction and design process are outlined in Section~\ref{sec:design} and Section~\ref{sec:casestudies}, respectively. Concluding remarks and directions for future work are provided in Section~\ref{sec:Conclusions}. 

\section{Preliminaries and System Dynamical Models} \label{section:dynamics}
In this section, we outline pertinent notation, and describe the dynamical model for the generators and the frequency-responsive DERs.

\subsection{Notation} 
The spaces of $N\times1$ real-valued and complex-valued vectors are denoted by $\mathbb{R}^N$ and $\mathbb{C}^N$, respectively. The matrix inverse is denoted by  $(\cdot)^\mathrm{-1}$, transpose by $(\cdot)^\mathrm{T}$, and $\mathrm{j} := \sqrt{-1}$. The magnitude of a complex scalar and cardinality of a set is denoted by $|\cdot|$. A diagonal matrix formed with diagonal entries composed of entries of vector $x$ is denoted by $\mathrm{diag}(x)$; and $\mathrm{diag}\{x,y\}$ denotes a diagonal matrix with entries of vectors $x$ and $y$ staked along the main diagonal. The $N \times 1$ vectors with all ones and all zeros are denoted by $1_{N}$ and $0_{N}$, respectively; and the $N \times N$ identity matrix is denoted by $I_N$. 

\subsection{Transmission Network Model}
We consider a classical power network model for the transmission grid, which is represented by a graph, where $\N$ is the set of buses, and $\E\subset \N\times\N$ is the set of transmission lines. A transmission line is denoted by $(\gen,\ell) \in \E$.  Partition the set $\N = \cD \cup \cG$, where $\G$ is the set of buses that are connected to conventional turbine-based generators (high inertia) and $\D$ is the set of buses that are connected to frequency-responsive DERs (or their aggregates). For notational and expositional convenience, we assume that no DERs are connected to generator buses, i.e., $\cD \cap \cG = \emptyset$. (This assumption can be easily relaxed at the risk of having to contend with burdensome notation.) The set of neighbors of bus $\gen$ is defined as $\N_\gen:= \{ \ell \in \N | \, (\gen,\ell) \in \E \}$. Transmission line $(\gen,\ell)$ is modeled as a series reactance $\mathrm{j} x_{\gen \ell} \in \mathbb{C} \setminus\{0\}$. The branch flows $P_{\gen \ell}$, $Q_{\gen \ell}$, are given by 
\begin{align} \label{eq.powerbalance}
\begin{split}
P_{\gen \ell} & = |V_\gen||V_\ell| x_{\gen \ell}^{-1} \sin (\theta_\gen-\theta_\ell), \\
Q_{\gen \ell} & = |V_\gen|^2 x_{\gen \ell}^{-1} - |V_\gen||V_\ell| x_{\gen \ell}^{-1} \cos(\theta_\gen-\theta_\ell),
\end{split}
\end{align}
where $|V_\ell|$ is the voltage magnitude and $\theta_\ell$ is the phase angle at the $\ell$ bus.  

\subsection{Synchronous-generator Dynamics}
\label{sec:gen}
Since we are interested in time scales pertaining to primary frequency response, we model the dynamics of angular position, frequency, and mechanical-power input for the generators in the network. In particular, for the $\gen \in \G$ generator, we adopt the following third-order dynamical model:
\begin{subequations}
\begin{align}
\dot\theta_\gen & = \omega_\gen - \omega_\mathrm{s},  \label{eq:gen_theta}\\
M_{\G,\gen} \dot\omega_\gen & =   P^{\mathrm{m}}_\gen  - D_{\G,\gen} (\omega_\gen - \omega_\mathrm{s}) + P_\gen - \sum_{\ell \in \N_\gen} P_{\gen \ell},\label{eq:gen_omega}\\  
\tau_{\gen} \dot P^{\mathrm{m}}_\gen &= - P^{\mathrm{m}}_\gen + P_\gen^\mathrm{r} - R_{\G,\gen}  (\omega_\gen - \omega_\mathrm{s}). \label{eq:gen_p} 
\end{align}
\end{subequations}
Above, $\theta_\gen$, $\omega_\gen$, and $P^{\mathrm{m}}_\gen$ are the dynamical states for rotor electrical angular position, generator frequency, and turbine mechanical power, respectively, for the $\gen$ generator, and $\omega_\mathrm s$ is the synchronous frequency. Furthermore, $M_{\cG,\gen}$ is the inertia constant, $D_{\cG,\gen}$ is the load-damping coefficient, $R_{\G,\gen}$ is the inverse of the frequency-power speed-droop regulation constant, $\tau_\gen$ is the turbine time constant, and $P^{\mathrm{r}}_\gen$ denotes its reference power setting (assumed to be constant since it derives from secondary control). Finally, $P_\gen$ is the injection at bus $\gen$ (negative, if we wish to model a constant power load). The above model is widely used for studying power-system dynamic phenomena at time scales pertaining to primary-frequency response~\cite{Apostolopoulou-2016,Dhople-2016}. Dynamics of automatic voltage regulators and power-system stabilizers are typically neglected for this regime, and the terminal voltage $|V_\gen|$ is fixed. For notational convenience, we define the following: 
\begin{align} 
P^\mathrm m&:= [P_1^\mathrm m, P_2^\mathrm m, \dots, P_{|\cG|}^\mathrm m]^\mathrm T, P^\mathrm{r} := [P_1^\mathrm r, P_2^\mathrm r, \dots, P_{|\cG|}^\mathrm r]^\mathrm T, \nonumber \\
M_{\cG} & := [M_{\cG, 1}, \dots, M_{\cG, |\cG|}]^\mathrm T, D_{\cG} := [D_{\cG, 1}, 
\dots, D_{\cG, |\cG|}]^\mathrm T, \nonumber \\
R_{\cG} & := [R_{\cG, 1}, \dots, R_{\cG, |\cG|}]^\mathrm T, \tau := [\tau_1, \tau_2, \dots, \tau_{|\cG|}]^\mathrm T \label{eq:machineparams}.
\end{align}

\subsection{Frequency-responsive DER Model} \label{sec.distnetmodel}
Assume the following model for the DERs connected to buses $d \in \mathcal D$:
\begin{subequations}
\begin{align} 
\dot \theta_d &= \omega_d - \omega_\mathrm s \label{eq:der_theta} \\
M_{\mathcal D, d} \dot \omega_d &= P_d  - \sum_{\ell \in \N_d} P_{d \ell} - D_{\mathcal D,d} (\omega_d - \omega_\mathrm s). \label{eq:der_omega}  
\end{align}
\end{subequations}
The droop coefficient $D_{\cD,d}$ establishes the frequency response of the DER at bus $d$, and the synthetic-inertia constant $M_{\mathcal D,d}$ determines the inertial response. If node $d$ is a regular load bus with no frequency-response DERs, then we simply set $D_{\cD,d} = M_{\cD,d} = 0$. Furthermore, $P_d$ denotes the net real-power injection into bus $d$ (constant, frequency-independent real-power loads are incorporated into $P_d$). We neglect DER capacity limits (to preserve analytical convenience), and DER internal-controller dynamics (since these would be executed at much faster time scales). The above model is appropriate for aggregations of DERs in a setting where the frequency at the feeder head (connected to the transmission network) percolates down to all buses in the feeder~\cite{donnelly2010frequency,Swaroop-2016}. For notational convenience, we define the following vectors:
\begin{align}
\begin{split}
M_{\cD} &:= [M_{\cD, 1}, \dots, M_{\cD, |\cD|}]^\mathrm T, \\
D_{\cD} &:= [D_{\cD, 1}, \dots, D_{\cD, |\cD|}]^\mathrm T. 
\end{split}
\end{align}

\section{Reduced Second-order Model and Accuracy} \label{sec:modelreduction}
In this section, we derive the reduced second-order model. 
\subsection{State-space Model} \label{section:steadystate}
The following discussion assumes that the electrical distances between geographically different parts of the power network are negligible, and therefore all the buses have the same frequency even during the transient~\cite{Ilic-2012}. Extensions to the case where this assumption may not hold (e.g., when the network has multiple balancing areas or is composed of weakly connected clusters) is part of ongoing effort. Assume the system initially operates at the steady-state equilibrium point with $\omega_\gen = \omega_d = \omega_\mathrm{s},\,\forall \gen \in \cG, d \in \cD$. Defining $\Delta \omega = \omega_\gen - \omega_\mathrm s = \omega_d - \omega_\mathrm s$, we get the following dynamics from~\eqref{eq:gen_omega} and~\eqref{eq:der_omega}
\begin{align}
M_{\G,\gen} \Delta \dot \omega & =   P^{\mathrm{m}}_\gen  - D_{\G,\gen} \Delta \omega + P_\gen - \sum_{\ell \in \N_\gen} P_{\gen \ell},\label{eq:commonfreqG}\\  
M_{\mathcal D, d} \Delta \dot \omega &= - D_{\mathcal D,d} \Delta \omega +P_d - \sum_{\ell \in \N_d} P_{d\ell}. \label{eq:commonfreqD} 
\end{align}
Summing~\eqref{eq:commonfreqG} over all $\gen \in \cG$, and~\eqref{eq:commonfreqD} over all $d \in \cD$, we get
\begin{equation}
M_{\mathrm{eff}} \Delta \dot \omega = 1_{|\cG|}^\mathrm T  P^{\mathrm{m}}  -D_{\mathrm{eff}} \Delta \omega + P_\mathrm{load}, \label{eq:commonfreq}
\end{equation}
where we define the \emph{effective inertia constant}, $M_\mathrm{eff}$, and \emph{effective damping constant}, $D_\mathrm{eff}$, as 
\begin{equation}\label{eq:eff}
M_{\mathrm{eff}} := 1_{|\cG|}^\mathrm T M_{\cG} + 1_{|\cD|}^\mathrm T M_{\cD}, 
D_{\mathrm{eff}} := 1_{|\cG|}^\mathrm T D_{\cG} + 1_{|\cD|}^\mathrm T D_{\cD},
\end{equation}
respectively. Going back to~\eqref{eq:commonfreq}, $P_{\mathrm{load}}$ is the total electrical load given by
\begin{align}
P_{\mathrm{load}} &:= \sum_{\gen \in \cG} \big(P_\gen - \sum_{\ell \in \N_\gen} P_{\gen \ell} \big) + \sum_{d \in \cD} \big(P_d - \sum_{\ell \in \N_d} P_{d\ell}\big) \nonumber \\
&= \sum_{\gen \in \cG} P_\gen + \sum_{d \in \cD} P_d.
\end{align}
The second equality follows from the fact that since we consider a lossless transmission network~\eqref{eq.powerbalance}: 
\begin{equation*}
\sum_{\gen \in \cG} \sum_{\ell \in \N_\gen} P_{\gen \ell} + \sum_{d \in \cD} \sum_{\ell \in \N_d} P_{d\ell} = 0.
\end{equation*}
Furthermore, collecting copies of~\eqref{eq:gen_p} $\forall \gen \in \cG$, we can write
\begin{equation}
\mathrm{diag}(\tau) \dot P^{\mathrm{m}} = - P^{\mathrm{m}} + P^\mathrm{r} - R_{\G} \Delta \omega. \label{eq:commonfreqP}
\end{equation}
We combine~\eqref{eq:commonfreq} and~\eqref{eq:commonfreqP} into the following standard state-space model:
\begin{equation} \label{eq:ss}
\dot x = A x + B u.
\end{equation}
The state vector and input, $x, u \in \mathbb{R}^{|\cG|+1}$, and system matrices, $A, B \in \mathbb{R}^{|\cG|+1 \times |\cG|+1}$ are given by
\begin{align}
x &= [\Delta \omega, (P^\mathrm m)^\mathrm T]^\mathrm T, \quad 
u = [P_\mathrm{load}, (P^\mathrm r)^\mathrm T]^\mathrm T, \\
A &= \left[
 \begin{array}{c c}
-D_{\mathrm{eff}}M_{\mathrm{eff}}^{-1}  & M_{\mathrm{eff}}^{-1} 1_{|\cG|}^\mathrm T \\
A_{R} & A_\tau
 \end{array} 
 \right], \,\,
B = \mathrm{diag}\{M_{\mathrm{eff}}^{-1}, -A_\tau \}, \nonumber
\end{align} 
where 
\begin{equation}
A_\tau = -\mathrm{diag}(\tau)^{-1}, \quad A_R = A_\tau R_\cG.
\end{equation}
With the state-space model in~\eqref{eq:ss} in place, we now develop a second-order model under the assumption that the values of the turbine-governor time constants are similar.

\subsection{Reduced Second-order Model}
Consider the following reduced second-order model to capture the frequency dynamics:
\begin{equation} \label{eq:ss_red}
\dot x_\mathrm{red} = A_\mathrm{red} x_\mathrm{red} + B_\mathrm{red} u_\mathrm{red}.
\end{equation}
The state vector and input, $x_\mathrm{red}, u_\mathrm{red} \in \mathbb{R}^2$, and system matrices, $A_\mathrm{red}, B_\mathrm{red} \in \mathbb{R}^{2\times2}$ are given by
\begin{align}
x_\mathrm{red} &= [\Delta \omega_\mathrm{red}, P^{\mathrm m}_\mathrm{red}]^\mathrm T , \quad u_\mathrm{red} = [P_\mathrm{load}, P^{\mathrm r}_\mathrm{red}]^\mathrm T, \\
A_\mathrm{red} &= \left[
 \begin{array}{c c}
-D_{\mathrm{eff}}M_{\mathrm{eff}}^{-1}  & M_{\mathrm{eff}}^{-1} \\
-R_{\cG,\mathrm{eff}} \overline\tau^{-1} & - \overline\tau^{-1}
 \end{array} 
 \right], \,\,
B_\mathrm{red} = \left[
 \begin{array}{c c}
M_{\mathrm{eff}}^{-1}  & 0 \\
0 & \overline\tau^{-1}
 \end{array} 
 \right] \nonumber
\end{align}
where $\overline \tau > 0$ is a model parameter (we comment more on this shortly and outline an optimization-based approach to determine it in Section~\ref{sec:pickingoverlinetau}), and 
\begin{equation}
P^{\mathrm r}_\mathrm{red} = 1_{|\cG|}^\mathrm T P^{\mathrm r}, \,\, R_{\cG,\mathrm{eff}} = 1_{|\cG|}^\mathrm T R_{\cG}.
\end{equation}
When $\tau_\gen = \tau_\ell,\, \forall \gen, \ell \in \cG$, it is straightforward to show that with the choice $\overline \tau = \tau_\gen$ and $\Delta \omega(0) = \Delta \omega_\mathrm{red}(0)$; $\Delta \omega(t) = \Delta \omega_\mathrm{red}(t), \,\, \forall \,t \geq 0$. Indeed, the reduced-order model above is conceptualized starting from this observation and under the assumption that in practice the turbine-governor time constants are similar in value~\cite{ramanujam2009power}. We show next that when the turbine time constants are not the same, the error in $\Delta \omega_\mathrm{red}(t)$ and $\Delta \omega(t)$ can be rigorously upper-bounded. 

\subsection{Accuracy of Reduced-order Model}
We derive an upper bound on the difference between $\Delta \omega_\mathrm{red}(t)$ and $\Delta \omega(t)$ for the general case when not all turbine time constants are equal. In order to accomplish this, we find it useful to define the following auxiliary dynamical system:
\begin{equation} \label{eq:ss_avg}
\dot{\overline x} = \overline A \overline x + \overline B u.
\end{equation}
The state vector, $\overline x \in \mathbb{R}^{|\cG|+1}$, and system matrices, $\overline A, \overline B \in \mathbb{R}^{|\cG|+1 \times |\cG|+1}$ are given by
\begin{align}
\overline x &= [\Delta \overline \omega, (\overline{P}^\mathrm m)^\mathrm T]^\mathrm T, \quad 
u = [P_\mathrm{load}, (P^\mathrm r)^\mathrm T]^\mathrm T, \\
\overline A &= \Gamma A =  \left[
 \begin{array}{c c}
-D_{\mathrm{eff}}M_{\mathrm{eff}}^{-1}  & M_{\mathrm{eff}}^{-1} 1_{|\cG|}^\mathrm T \\
\overline{A}_{R} & \overline{A}_\tau
 \end{array} 
 \right], \,\, \\
\overline B &= \Gamma B = \mathrm{diag}\{M_{\mathrm{eff}}^{-1}, -\overline A_\tau \}, \nonumber
\end{align} 
where 
\begin{align}
\Gamma &= \mathrm{diag}\{1,\overline{\tau}^{-1}\mathrm{diag}(\tau)\}, \\
\overline A_\tau &= -\overline{\tau}^{-1} I_{|\cG|}, \quad \overline A_R = \overline A_\tau R_\cG.
\end{align}
Suppose the initial conditions for the system~\eqref{eq:ss_avg} are picked to match those of~\eqref{eq:ss}, i.e., $\overline x(0) = x(0)$. The reduced-order model in~\eqref{eq:ss_red} can be derived from the one in~\eqref{eq:ss_avg} with the choice $P^{\mathrm m}_{\mathrm{red}} = 1_{|\cG|}^\mathrm{T} \overline{P}^\mathrm{m}$ and $P^{\mathrm r}_{\mathrm{red}} = 1_{|\cG|}^\mathrm{T} P^\mathrm{r}$. Furthermore, with the initialization $\Delta \omega_\mathrm{red}(0) = \Delta \overline \omega(0)$, and $P^{\mathrm m}_{\mathrm{red}}(0) = 1_{|\cG|}^\mathrm{T} \overline{P}^\mathrm{m}(0)$, it follows that $\Delta \omega_\mathrm{red}(t) = \Delta \overline \omega(t),\,\,\forall t \geq 0$. 

Since $\Delta \omega_\mathrm{red}(t) = \Delta \overline \omega(t),\,\,\forall t \geq 0$, and since the system~\eqref{eq:ss_avg} has the same dimension as the original system~\eqref{eq:ss}, it is algebraically and analytically convenient to study~\eqref{eq:ss_avg} (instead of~\eqref{eq:ss_red} directly) as compared to~\eqref{eq:ss}. We note that if not all the entries of $\tau$ are identical, then the trajectories generated by~\eqref{eq:ss_avg} and~\eqref{eq:ss} do not match. Nonetheless, if the eigenvalues of $A$ and those of $\Gamma A$ are close, then we would expect $x(t)$ and $\overline x(t)$ to be close, and hence $\Delta \omega(t)$ and $\Delta \omega_\mathrm{red}(t)$ to be close. We utilize the $2$-norm of the matrix
\begin{equation} \label{eq:definitions}
E := \overline A - A = (\Gamma - I_{|\cG|+1}) A 
\end{equation}
as a measure of closeness of eigenvalues of $A$ and $\overline A$, and building on this, we derive an upper bound to the difference between $\Delta \omega(t)$ and $\Delta \omega_\mathrm{red}(t)$ that is proportional to $\|E\|_2$. With this final goal in mind, we will find the following lemma useful.
\begin{lemma}
Suppose the matrix $A$ in~\eqref{eq:ss} is Hurwitz and diagonalizable. There exists a $\delta > 0$, such that if $\|E\|_2 = \|(\Gamma - I_{|\cG|+1})A\|_2 < \delta$, then the matrix $\Gamma A$ is Hurwitz.
\end{lemma}

\noindent \emph{Proof.} Diagonalize matrix $A$ as $A = P \Lambda P^{-1}$. By the theory of perturbation bounds for eigenvalues and eigenvectors~\cite{karow2014perturbation}, and the definitions in~\eqref{eq:definitions}, we have that
\begin{equation*}\label{eq:taylor-A-prime}
\Gamma A = A + E = \left(P+ g(E) \right) (\Lambda + h(E)) \left(P+ g(E) \right)^{-1},
\end{equation*}
where $g(E) \in \mathbb{C}^{|\cG| +1 \times |\cG| +1}$, and $h(E) \in \mathbb{C}^{|\cG| +1 \times |\cG| +1}$ is a diagonal matrix. Furthermore, $g(E) = h(E)  = O(\|E\|_2)$, which implies that 
\begin{equation} \label{eq:prelim1}
\|g(E)\|_2 = \|h(E)\|_2 \to 0 \,\,\,\, \text{as}\,\,\,\,\|E\|_2 \to 0.
\end{equation}
From~\eqref{eq:prelim1}, we can conclude that there exists a sufficiently small $\delta >0$ such that when $\|E\|_2 = \|(\Gamma - I_{|\cG|+1})A\|_2 < \delta$, the eigenvalues of $\overline A = \Gamma A$, i.e., the diagonal entries of the diagonal matrix $\Lambda + h(E)$, have strictly negative real parts.~\qedblack \\

Leveraging the result of Lemma~1, we now bound the error between $\Delta \omega_\mathrm{red}(t)$ and $\Delta \omega(t)$. \\

\begin{theorem} \label{thm:main}
Consider the dynamical system~\eqref{eq:ss} (with a matrix $A$ that is diagonalizable and Hurwitz) and the reduced-order counterpart~\eqref{eq:ss_red}. Suppose the initial conditions for the two dynamical systems at time $t = 0$ are such that $\Delta \omega_\mathrm{red}(0) = \Delta \omega(0)$, and $P^{\mathrm{m}}_{\mathrm{red}}(0) = 1_{|\cG|}^{\mathrm{T}} P^{\mathrm m}(0)$. There exist $\delta,k, \lambda > 0$, such that if $\|(\Gamma - I_{|\cG|+1})A)\|_2 < \delta$, 
we get that $\forall t \geq 0$,
\begin{equation} \label{eq:frequencybound}
|\Delta \omega(t) - \Delta \omega_\mathrm{red}(t)| < \delta \frac{k}{\lambda} \sup_{0\leq s\leq t} \left(\| x(s)\|_2 + \|A^{-1}B u(s)\|_2\right).
\end{equation}
\end{theorem}

\noindent\emph{Proof.} Consider the dynamics of $\Delta x(t) := \overline x(t) - x(t)$, which, given the models in~\eqref{eq:ss},~\eqref{eq:ss_avg}, and the definitions in~\eqref{eq:definitions} can be expressed as
\begin{align} \label{eq:delta_x}
\Delta \dot x &= \Gamma A \Delta x + (\Gamma - I_{|\cG|+1}) \dot x, \,\, \Delta x(0) = 0_{|\cG|+1}.
\end{align}
With a Hurwitz matrix $\Gamma A$ (see Lemma~1), and treating $\dot x$ as an exogenous input to the system in~\eqref{eq:delta_x}, we can write its solution as 
\begin{equation} \label{eq:delta_x_sol}
\Delta x(t) = \int_{s = 0}^{t} \mathrm{e}^{\Gamma A(t-s)}(\Gamma - I_{|\cG|+1}) \dot{x}(s) ds.
\end{equation}
Since $\Gamma A$ is Hurwitz, there exist $k, \lambda >0$ such that we can bound~\cite{Khalil_Book02}
\begin{equation} \label{eq:matrixscalarbound} 
\|\mathrm{e}^{\Gamma A(t-s)}\|_2 \leq k \mathrm{e}^{-\lambda (t-s)}, \, \, \forall  \,\, 0\leq s \leq t.
\end{equation}
From~\eqref{eq:delta_x_sol} and~\eqref{eq:matrixscalarbound}, we can write  
\begin{align} \label{eq:prelim3}
\|\Delta x(t)\|_2 &\leq \int_{s = 0}^{t} k \mathrm{e}^{-\lambda (t-s)} \nonumber \\ 
& \qquad \quad  \cdot \| (\Gamma - I_{|\cG|+1}) (Ax(s) + Bu(s))\|_2 ds \nonumber \\
&\leq \frac{k}{\lambda} \|(\Gamma - I_{|\cG|+1})A\|_2 \nonumber \\
& \qquad \quad  \cdot \sup_{0\leq s\leq t} (\| x(s)\|_2 + \|A^{-1}Bu(s)\|_2). 
\end{align}
Recognizing that
\begin{align*}
|\Delta \overline{\omega}(t) - \Delta \omega(t)| &= |\Delta \omega_\mathrm{red}(t) - \Delta \omega(t)| \leq \|\Delta x(t)\|_2,
\end{align*}
and under the constraint $\|(\Gamma - I_{|\cG|+1})A)\|_2 < \delta$, we get the bound in~\eqref{eq:frequencybound}.~\qedblack 

\subsection{Selecting an Appropriate $\overline{\tau}$} \label{sec:pickingoverlinetau}
Given the bound in Theorem 1 that depends on $\|(\Gamma - I_{\|\cG\|+1})A\|_2$, evidently, a good choice for $\overline \tau$ would be:
\begin{equation} \label{eq:overline1}
\overline \tau = \mathrm{arg}\min_{\widehat \tau\geq 0} \|(\Gamma(\widehat \tau) - I_{|\cG|+1})A \|_2, 
\end{equation}
where $\Gamma(\widehat \tau) := \widehat{\tau}^{-1}\mathrm{diag}\{\widehat \tau,\mathrm{diag}(\tau)\}$. Serendipitously, we find that 
\begin{equation*}
\|(\Gamma(\widehat \tau) - I_{|\cG|+1})A\|_2 = \|(\widetilde \Gamma(\widehat \tau) - I_{|\cG|}) \widetilde A\|_2,
\end{equation*}
where 
\begin{equation}
\widetilde{\Gamma}(\widehat \tau) := \widehat{\tau}^{-1} \mathrm{diag}(\tau), \quad \widetilde A := [A_R \, \, A_{\tau}],
\end{equation}
with $A_\tau = -\mathrm{diag}(\tau)^{-1}$ and $A_R = A_\tau R_\cG$ (see~\eqref{eq:ss}). This is because the first row of the matrix $(\Gamma(\widehat \tau) - I_{|\cG|+1})A$ has all entries equal to $0$. Therefore, in lieu of solving~\eqref{eq:overline1}, we solve instead 
\begin{equation} \label{eq:overline2}
\overline \tau = \mathrm{arg}\min_{\widehat \tau \geq 0} \|(\widetilde \Gamma(\widehat \tau) - I_{|\cG|})\widetilde A \|_2.
\end{equation}
This is an important point to emphasize since the matrix $\widetilde A$ does not depend on the effective damping and inertia constants ($D_\mathrm{eff}$ and $M_\mathrm{eff}$): the terms that we wish to design to engineer the primary-frequency and inertial response of the system. With the choice~\eqref{eq:overline2}, we see that the reduced-order model in~\eqref{eq:ss_red} is fully specified, and we can move on to the design process leveraging the analytical simplicity afforded by the second-order system.

\section{Designing Inertia and Damping Coefficients} \label{sec:design}
In order to tune parameters $M_\mathrm{eff}$ and $D_\mathrm{eff}$ in the combined transmission and distribution systems modeled by~\eqref{eq:ss} to achieve desired transient characteristics in system frequency deviations, we make use of the reduced second-order system in~\eqref{eq:ss_red}.  Particularly, we consider the $s$-domain transfer function from $P_\mathrm{load}$ to $\Delta \omega_{\mathrm{red}}$, which was shown to approximate the actual frequency deviation $\Delta \omega$ in Theorem~\ref{thm:main}.  This input-output relationship is easily obtained from~\eqref{eq:ss_red} as
\begin{equation}\label{eq:tf}
\frac{\Delta \omega_\mathrm{red} (s)}{P_\mathrm{load}(s)} = \frac{k (s+ a)}{s^2 + 2 \zeta \omega_{\mathrm{n}} s + \omega_{\mathrm{n}}^2} =: H(s),
\end{equation}
where 
\begin{align} \label{eq:omeganzeta}
k &= M_{\mathrm{eff}}^{-1}, \quad a = \overline \tau^{-1}, \\ 
\omega_{\mathrm{n}} &= \sqrt{\frac{R_{\cG, \mathrm{eff}} + D_{\mathrm{eff}}}{\overline\tau M_{\mathrm{eff}}}} , \quad \zeta = \frac{1}{2} \frac{M_\mathrm{eff} + \overline \tau D_\mathrm{eff}}{\sqrt{\overline \tau M_\mathrm{eff} (R_{\cG,\mathrm{eff}} + D_\mathrm{eff} )}}. \nonumber
\end{align} 

\begin{figure}[b]
\centering
\includegraphics[scale=1]{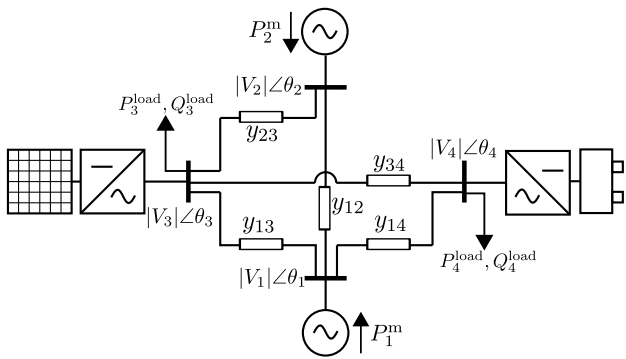}
\caption{One-line diagram of test case. Synchronous generators are at buses $\cG = \{1,2\}$, and frequency-responsive DERs are at buses $\cD = \{3,4\}$. Frequency-independent loads at buses $3$ and $4$ are denoted by $(P_3^\mathrm{load},Q_3^\mathrm{load})$ and $(P_4^\mathrm{load},Q_4^\mathrm{load})$.} \label{fig:testcase}
\end{figure}

\subsection{Steady-state Frequency Regulation}
\label{sec:ss_freq}
Suppose the specifications call for a steady-state regulation of $R_{\mathrm{reg}}$ for primary-frequency response. In particular, 
\begin{equation}
R_\mathrm{reg} = \frac{\Delta P_\mathrm{load}}{\Delta \omega^\star}, 
\end{equation}
where $\Delta P_\mathrm{load}$ is the (step) change in load from the pre-disturbance synchronous-steady-state equilibrium and $\Delta \omega^\star = \lim_{t \to \infty} \Delta \omega(t)$ is the permissible steady-state deviation in frequency from~$\omega_\mathrm s$. From~\eqref{eq:tf}, we get 
\begin{equation}
R_\mathrm{reg} = \lim_{s\to 0} H(s)^{-1} =  \frac{\omega^2_\mathrm{n}}{ka} = D_\mathrm{eff} + R_{\cG,\mathrm{eff}}.  
\end{equation}
Given a specification on $R_\mathrm{reg}$, and assuming that the generator damping coefficients are specified (collected in the vector $D_{\cG}$) the DER-side damping coefficients (i.e., entries of the vector $D_{\cD}$) should be picked to satisfy
\begin{equation} \label{eq.Dconstraint}
1_{|\cD|}^\mathrm T D_{\cD} = R_\mathrm{reg} - R_{\cG,\mathrm{eff}} - 1_{|\cG|}^\mathrm T D_{\cG}.
\end{equation}

Notice that~\eqref{eq.Dconstraint} establishes a constraint on the sum of DER damping coefficients. There are many possibilities to disaggregate this sum into individual values $D_{\cD,d}$. Let us suppose that we are interested in ensuring power sharing in proportion to ratings. In particular, in the post-disturbance equilibrium, we want the ratio of the change in real-power output from DER $d$ to its real-power rated value $P^\mathrm{rated}_d$ to be the same for all DERs. This can be ensured with the choice
\begin{equation}\label{eq:Dchoice}
\frac{D_{\cD,k}}{D_{\cD,\ell}} = \frac{P^\mathrm{rated}_k}{P^\mathrm{rated}_\ell}, \quad k, \ell \in \cD.
\end{equation}

\subsection{Transient Frequency Dynamics}
With $D_\mathrm{eff}$ determined to meet the steady-state frequency regulation requirement, the only tunable parameter remaining in~\eqref{eq:tf} is $M_\mathrm{eff}$. We can tune this parameter to engineer the desired transient performance. Given the transfer function~\eqref{eq:tf}, we can readily solve for $M_\mathrm{eff}$ that guarantees a prescribed value of damping, $\zeta$, or natural frequency, $\omega_\mathrm n$.

Once $M_\mathrm{eff}$ has been determined according to the desired transient performance of~\eqref{eq:tf}, we design $M_{\cD,d}$ for individual DERs $d \in \cD$ with a design philosophy that is similar to how the $D_{\cD,d}$ terms were obtained from $D_\mathrm{eff}$ in Section~\ref{sec:ss_freq}.  
To this end, suppose that we are interested in ensuring power sharing that is proportional to individual DER power ratings. At \emph{any time} $t \geq 0$, we want the ratio of the change in active-power output from DER $d$ for inertial response to its power rating $P^\mathrm{rated}_d$ to be the same for all DERs. This can be ensured with the choice
\begin{equation}\label{eq:Mchoice}
\frac{M_{\cD,k}}{M_{\cD,\ell}} = \frac{P^\mathrm{rated}_k}{P^\mathrm{rated}_\ell}, \quad \forall k, \ell \in \cD.
\end{equation}

\section{Numerical Simulation Results} \label{sec:casestudies}
Consider the 4-bus transmission system shown in Fig.~\ref{fig:testcase} with synchronous generators at buses $\cG = \{1, 2\}$ and frequency-responsive DERs at $\D = \{3,4\}$. The model parameters and power-flow states corresponding to the pre-disturbance steady state are listed in the appendix. Unless otherwise specified, voltage magnitudes are in per unit (pu) with a $4.8$ [$\mathrm{kV}$] base, and power values are also in pu with a $23$ [$\mathrm{MVA}$] base. To validate the second-order model in~\eqref{eq:ss_red} and design process in Section~\ref{sec:design}, pertinent time-domain simulation results are compared with a differential algebraic equation (DAE) model simulated in Power System Toolbox (PST)~\cite{chow2000power}. In addition to the model introduced in Section~\ref{sec:gen}, the PST DAE model also considers lossy lines, voltage-regulator dynamics, and a detailed two-axis machine model.  Furthermore, DER-connected nodes are represented as $PQ$ buses in the power-flow solution, and DERs are modeled as frequency-sensitive negative loads. (This conforms to the fact that state-of-art DERs are grid following devices.) We will find that the simulation results validate the modeling assumptions in deriving the state-space model in~\eqref{eq:ss} and the ensuing model-reduction method~\eqref{eq:ss_red}.  

\subsection{Accuracy of the Reduced-order Model}
Accuracy of the reduced second-order model is established with time-domain simulation results (for two choices of $\overline \tau$ including the one suggested in~\eqref{eq:overline2}) and examining the poles and zeros of the load-step to frequency-deviation transfer function of the original and reduced-order models.

\subsubsection{Choice of $\overline \tau$} 
\begin{figure} 
\centering
\includegraphics[width=1\linewidth]{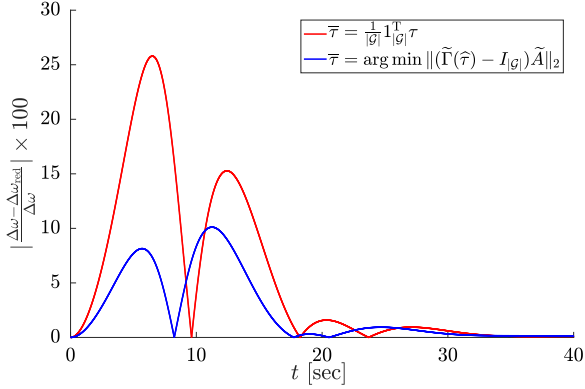} 
\caption{Error in reduced-order frequency-deviation dynamics for two different choices of $\overline \tau$.}
\label{fig:overlinetaufig}
\end{figure}

In Fig.~\ref{fig:overlinetaufig}, we plot the relative error between the electrical frequency deviation (from synchronous frequency) from the model~\eqref{eq:ss}, $\Delta \omega(t)$ and the frequency deviation resulting from the reduced-order model, $\Delta \omega_\mathrm{red}(t)$. The observed frequency deviations result from a $0.02$ [$\mathrm{MW}$] step change in the load, $P_3^\mathrm{load}$, at bus $3$. The two trajectories shown in Fig.~\ref{fig:overlinetaufig} correspond to cases for which the reduced-order system in~\eqref{eq:ss_red} is simulated with $\overline \tau$ picked as: i)~the average of the turbine time constants of generators $1$ and $2$ (red trace), and ii)~the solution to~\eqref{eq:overline2} (blue trace).  Clearly, the choice of $\overline \tau$ in~\eqref{eq:overline2} serves as a more suitable proxy for the turbine time constant of the aggregated governor dynamics. 

\subsubsection{Transfer Functions}
\begin{figure}
\includegraphics[scale = 0.5]{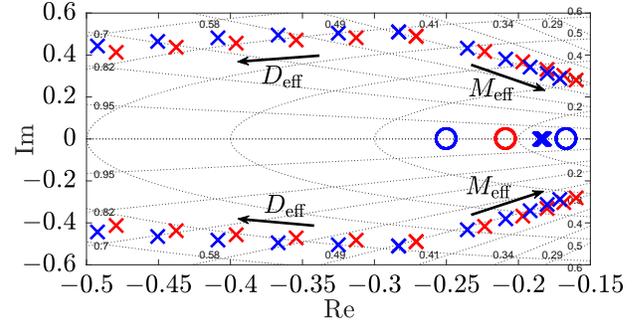}
\caption{Poles and zeros of the transfer functions from load disturbance to frequency deviation for the original model~\eqref{eq:ss} (blue), and the reduced second-order model~\eqref{eq:ss_red} (red) as $D_\mathrm{eff}$ and $M_\mathrm{eff}$ are varied.}
\label{fig:pzplot}
\end{figure}

In Fig.~\ref{fig:pzplot}, we plot the poles and zeros of the transfer functions from load disturbance to frequency deviation for the original model~\eqref{eq:ss} in blue, and the reduced second-order model~\eqref{eq:ss_red} in red (in which case, we refer to the transfer function in~\eqref{eq:tf}). The following observations are in order:
\begin{itemize}
 \renewcommand{\labelitemi}{\tiny$\blacksquare$}
 \item  Increasing $D_\mathrm{eff}$ (with $M_\mathrm{eff}$ held constant) perturbs the complex-valued poles predominantly along trajectories associated with constant natural frequency.
 \item Increasing $M_\mathrm{eff}$ (with $D_\mathrm{eff}$ held constant) perturbs the complex-valued poles predominantly along trajectories associated with constant damping ratio. 
\item Zeros are independent of $D_\mathrm{eff}$ and $M_\mathrm{eff}$ values in both models (this is evident for the reduced-order model from~\eqref{eq:omeganzeta}). Furthermore, real-valued poles of the original model (predominantly attributable to the governor dynamics) are not perturbed significantly.
\item Most importantly, complex-valued poles corresponding to the reduced second-order model are close to those of original model with $\overline \tau$ chosen by solving~\eqref{eq:overline2} over a wide range of $D_\mathrm{eff}$ and $M_\mathrm{eff}$. 
\end{itemize}

\begin{figure}
\includegraphics[width=\linewidth]{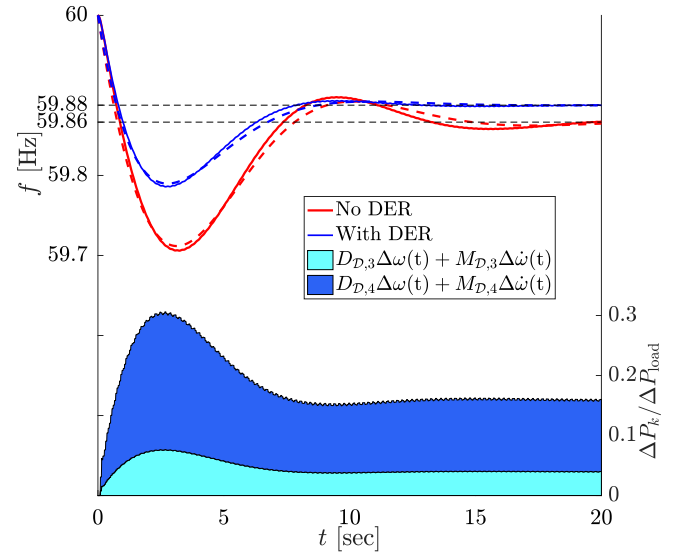} 
\caption{(top) Transient frequency dynamics and steady-state frequency regulation is improved with DERs. Dashed (red and blue) lines are generated by simulating the second-order model, and solid lines are obtained from the PST simulation. Analytically computed steady-state frequencies are shown in dashed black lines. (We design for $R_\mathrm{reg} = 0.4644$ and $\zeta = 0.7$.) (bot) Power outputs of DERs illustrate proportional sharing for inertial and primary-frequency response. ($P^\mathrm{rated}_3/P^\mathrm{rated}_4 = 1/3$.)}
\label{fig:timedomain}
\end{figure}

\subsection{Achieving Desired Performance Specifications}
With the reduced-order model validated, we apply the ideas outlined in Section~\ref{sec:design} to design the DER synthetic-inertia and droop-control parameters. To this end, consider the trajectories in the top pane of Fig.~\ref{fig:timedomain}. Dashed traces correspond to simulations from the reduced-order model, while solid traces correspond to those obtained from PST.  For both, at time $t=0$, a step increase of $0.02$ [$\mathrm{MW}$] is applied to the load, $P_3^\mathrm{load}$, at bus $3$. Trajectories in red correspond to the case where the DERs do not participate in frequency response (i.e., $D_{\cD,d} = M_{\cD,d} = 0,\,d \in \{3,4\}$.) Compared to this base case, we show the frequency response (due to the same load increase) with the DER synthetic-inertia and droop-control parameters designed for a frequency regulation $R_\mathrm{reg} = 0.4644$ and $\zeta = 0.7$ (traces in blue). Specifically, requiring $R_\mathrm{reg} = 0.4644$ sets $D_\mathrm{eff} = 0.0738$, and requiring $\zeta = 0.7$ fixes $M_\mathrm{eff} = 0.0111$.  Notice that the chosen value of $\zeta$ yields a damped response with a lower frequency nadir. Furthermore, the steady-state frequency deviation from synchronous frequency is also significantly reduced. Trajectories generated from the second-order model match those obtained from the PST simulation. (This is noteworthy, given that the PST model considers lossy lines, voltage-regulator dynamics, and a detailed two-axis machine model.) 

With $D_\mathrm{eff}$ and $M_\mathrm{eff}$ chosen, the damping-constant and synthetic-inertia values for individual DERs are picked based on~\eqref{eq:Dchoice} and~\eqref{eq:Mchoice}, respectively. Power outputs of the two DERs are shown in the bottom pane of Fig.~\ref{fig:timedomain}. Notice that the DERs indeed share the load increase in proportion to their power ratings ($P^\mathrm{rated}_3/P^\mathrm{rated}_4 = 1/3$) across time scales pertinent to inertial and primary-frequency response.

\section{Concluding Remarks \& Directions for Future Work} \label{sec:Conclusions}
We proposed a framework to engineer synthetic-inertia and droop-control parameters for distributed energy resources (DERs) so that the system frequency in a network composed of DERs and synchronous generators conforms to prescribed transient and steady-state performance specifications. Our approach was developed by formulating a lumped-parameter reduced second-order model for frequency dynamics. This allowed us to systematically design the DER inertial and droop-control coefficients leveraging classical frequency-domain response characteristics of second-order systems. As part of ongoing efforts, we are extending the method to cover networks composed of multiple balancing areas. 

\section*{Appendix}
The synchronous frequency, $\omega_{\mathrm s} = 2 \pi 60 \,[ \mathrm{rad}\,\mathrm {sec^{-1}}]$. All values are reported in per unit unless otherwise noted. The generator damping coefficients are: $D_{\cG, 1} = D_{\cG, 2} = 0.0434$, inertia constants are: $M_{\cG, 1} = M_{\cG, 2} = 0.1302 \, [\mathrm{sec}]$, droop coefficients are: $R_{\cG, 1} = 0.217$ and $R_{\cG, 2} = 0.0868$, turbine time constants are $\tau_1 = 4\,[\mathrm{sec}]$ and $\tau_2 = 10\,[\mathrm{sec}]$, reference power values are $P_1^\mathrm{r} = 0.0109, P_2^\mathrm{r} = 0.0043, Q_2^\mathrm{r} = 0.0061$ and $Q_2^\mathrm{r} = 0$. Frequency-independent loads at buses $3$ and $4$ are denoted by $(P_3^\mathrm{load} = 0.0217 ,Q_3^\mathrm{load} = 0.0065)$ and $(P_4^\mathrm{load} = 0.0087,Q_4^\mathrm{load} =0)$. The rated power outputs of the DERs are $P_3^\mathrm{rated} = 0.25$ and $P_4^\mathrm{rated} =0.75$. The transmission line parameters are given by $y_{12} = 0.5+\mathrm j 10 , y_{13} = 0.5+\mathrm j 5, y_{14} = 1+\mathrm j 5, y_{23} = 0.5+\mathrm j 5$, and $y_{34} = 1+\mathrm j 5$. 

\bibliographystyle{IEEEtran}
\bibliography{references}

\begin{thebibliography}{10}
\providecommand{\url}[1]{#1}
\csname url@samestyle\endcsname
\providecommand{\newblock}{\relax}
\providecommand{\bibinfo}[2]{#2}
\providecommand{\BIBentrySTDinterwordspacing}{\spaceskip=0pt\relax}
\providecommand{\BIBentryALTinterwordstretchfactor}{4}
\providecommand{\BIBentryALTinterwordspacing}{\spaceskip=\fontdimen2\font plus
\BIBentryALTinterwordstretchfactor\fontdimen3\font minus
  \fontdimen4\font\relax}
\providecommand{\BIBforeignlanguage}[2]{{%
\expandafter\ifx\csname l@#1\endcsname\relax
\typeout{** WARNING: IEEEtran.bst: No hyphenation pattern has been}%
\typeout{** loaded for the language `#1'. Using the pattern for}%
\typeout{** the default language instead.}%
\else
\language=\csname l@#1\endcsname
\fi
#2}}
\providecommand{\BIBdecl}{\relax}
\BIBdecl

\bibitem{Taylor-2016}
J.~A. Taylor, S.~V. Dhople, and D.~S. Callaway, ``Power systems without fuel,''
  \emph{Renewable and Sustainable Energy Reviews}, vol.~57, pp. 1322--1336,
  2016.

\bibitem{Sullivan-2014}
J.~O'Sullivan, A.~Rogers, D.~Flynn, P.~Smith, A.~Mullane, and M.~O'Malley,
  ``{Studying the Maximum Instantaneous Non-Synchronous Generation in an Island
  System---Frequency Stability Challenges in Ireland},'' \emph{IEEE
  Transactions on Power Systems}, vol.~29, no.~6, pp. 2943--2951, November
  2014.

\bibitem{Ela-2014}
E.~Ela, V.~Gevorgian, A.~Tuohy, B.~Kirby, M.~Milligan, and M.~O'Malley,
  ``{Market Designs for the Primary Frequency Response Ancillary Service---Part
  I: Motivation and Design},'' \emph{IEEE Transactions on Power Systems},
  vol.~29, no.~1, pp. 421--431, January 2014.

\bibitem{Gevorgian-2015}
V.~Gevorgian, Y.~Zhang, and E.~Ela, ``Investigating the impacts of wind
  generation participation in interconnection frequency response,'' \emph{IEEE
  Transactions on Sustainable Energy}, vol.~6, no.~3, pp. 1004--1012, July
  2015.

\bibitem{chow2013power}
J.~Chow, \emph{Power System Coherency and Model Reduction}, ser. Power
  Electronics and Power Systems.\hskip 1em plus 0.5em minus 0.4em\relax
  Springer, New York, 2013.

\bibitem{Apostolopoulou-2016}
D.~Apostolopoulou, P.~W. Sauer, and A.~D. Dom\'{i}nguez-Garc\'{i}a, ``Balancing
  authority area model and its application to the design of adaptive agc
  systems,'' \emph{IEEE Transactions on Power Systems}, vol.~31, no.~5, pp.
  3756--3764, September 2016.

\bibitem{Ghosh-2016}
S.~Ghosh, S.~Kamalasadan, N.~Senroy, and J.~Enslin, ``{Doubly Fed Induction
  Generator (DFIG)-Based Wind Farm Control Framework for Primary Frequency and
  Inertial Response Application},'' \emph{IEEE Transactions on Power Systems},
  vol.~31, no.~3, pp. 1861--1871, May 2016.

\bibitem{Zhao-2016}
J.~Zhao, X.~Lyu, Y.~Fu, X.~Hu, and F.~Li, ``Coordinated microgrid frequency
  regulation based on dfig variable coefficient using virtual inertia and
  primary frequency control,'' \emph{IEEE Transactions on Energy Conversion},
  vol.~31, no.~3, pp. 833--845, September 2016.

\bibitem{Vyver-2016}
J.~V. de~Vyver, J.~D. M.~D. Kooning, B.~Meersman, L.~Vandevelde, and T.~L.
  Vandoorn, ``Droop control as an alternative inertial response strategy for
  the synthetic inertia on wind turbines,'' \emph{IEEE Transactions on Power
  Systems}, vol.~31, no.~2, pp. 1129--1138, March 2016.

\bibitem{Baldick-2014}
H.~Ch\'{a}vez, R.~Baldick, and S.~Sharma, ``Governor rate-constrained opf for
  primary frequency control adequacy,'' \emph{IEEE Transactions on Power
  Systems}, vol.~29, no.~3, pp. 1473--1480, May 2014.

\bibitem{Fei-2015}
F.~Teng, M.~Aunedi, D.~Pudjianto, and G.~Strbac, ``{Benefits of Demand-Side
  Response in Providing Frequency Response Service in the Future GB Power
  System},'' \emph{Frontiers in Energy Research}, vol.~3, p.~36, 2015.

\bibitem{Teng-2016}
F.~Teng, V.~Trovato, and G.~Strbac, ``Stochastic scheduling with
  inertia-dependent fast frequency response requirements,'' \emph{IEEE
  Transactions on Power Systems}, vol.~31, no.~2, pp. 1557--1566, March 2016.

\bibitem{Borsche-2015}
T.~S. Borsche, T.~Liu, and D.~J. Hill, ``Effects of rotational inertia on power
  system damping and frequency transients,'' in \emph{IEEE Conference on
  Decision and Control}, December 2015, pp. 5940--5946.

\bibitem{Poolla-2016}
B.~K. Poolla, S.~Bolognani, and F.~Dorfler, ``Placing rotational inertia in
  power grids,'' in \emph{2016 American Control Conference}, July 2016, pp.
  2314--2320.

\bibitem{Swaroop-2016}
S.~S. Guggilam, C.~Zhao, E.~Dall'Anese, Y.~C. Chen, and S.~V. Dhople,
  ``{Primary Frequency Response with Aggregated DERs},'' in \emph{American
  Control Conference}, 2017, to appear.

\bibitem{Dhople-2016}
H.~Choi, P.~J. Seiler, and S.~V. Dhople, ``Propagating uncertainty in
  power-system dae models with semidefinite programming,'' \emph{IEEE
  Transactions on Power Systems}, 2016, To appear.

\bibitem{donnelly2010frequency}
M.~Donnelly, D.~Harvey, R.~Munson, and D.~Trudnowski, ``Frequency and stability
  control using decentralized intelligent loads: Benefits and pitfalls,'' in
  \emph{Proceedings of IEEE Power and Energy Society General Meeting},
  Minneapolis, MN, USA, 2010, pp. 1--6.

\bibitem{Ilic-2012}
M.~D. Ili{\'{c}} and Q.~Liu, \emph{Toward Sensing, Communications and Control
  Architectures for Frequency Regulation in Systems with Highly Variable
  Resources}.\hskip 1em plus 0.5em minus 0.4em\relax New York, NY: Springer New
  York, 2012, pp. 3--33.

\bibitem{ramanujam2009power}
R.~Ramanujam, \emph{Power System Dynamics: Analysis and Simulation}.\hskip 1em
  plus 0.5em minus 0.4em\relax PHI Learning Pvt. Ltd., 2009.

\bibitem{karow2014perturbation}
M.~Karow and D.~Kressner, ``On a perturbation bound for invariant subspaces of
  matrices,'' \emph{SIAM Journal on Matrix Analysis and Applications}, vol.~35,
  no.~2, pp. 599--618, 2014.

\bibitem{Khalil_Book02}
H.~Khalil, \emph{Nonlinear Systems}, 3rd~ed.\hskip 1em plus 0.5em minus
  0.4em\relax Upper Saddle River, NJ: Prentice Hall, 2002.

\bibitem{chow2000power}
J.~Chow and G.~Rogers, \emph{Power System Toolbox}.\hskip 1em plus 0.5em minus
  0.4em\relax Cherry Tree Scientific Software, 2000.

\end{thebibliography}
\end{document}